\documentclass{emulateapj}

\usepackage{amsmath}
\usepackage{apjfonts}
\usepackage{amssymb}
\usepackage{mathrsfs}
\usepackage{color}
\usepackage{natbib}
\shorttitle{Tidal Shock Breakout}
\shortauthors{Guillochon, Ramirez-Ruiz, Rosswog \& Kasen}

\begin{document}

\title{Three-Dimensional Simulations of Tidally Disrupted Solar-Type
  Stars and the Observational Signatures of Shock Breakout}

\author{James Guillochon\altaffilmark{1}, Enrico
  Ramirez-Ruiz\altaffilmark{1}, Stephan Rosswog\altaffilmark{2}, and Daniel Kasen\altaffilmark{1,3}}

\altaffiltext{1}{Department of Astronomy and Astrophysics, University
  of California, Santa Cruz, CA 95064, USA}
\altaffiltext{2}{School of
  Engineering and Science, Jacobs University Bremen, Campus Ring 1,
  28759 Bremen, Germany}
\altaffiltext{3}{Hubble Fellow}

\begin{abstract}
We describe a three-dimensional simulation of a $1 M_{\odot}$
solar-type star approaching a $10^{6} M_{\odot}$ black hole on a
parabolic orbit with a pericenter distance well within the tidal
radius. While falling towards the black hole, the star is not only
stretched along the orbital direction but even more severely
compressed at right angles to the orbit. The overbearing degree of
compression achieved shortly after pericenter leads to the production
of strong shocks which largely homogenize the temperature profile of
the star, resulting in surface temperatures comparable to the initial
temperature of the star's core. This phenomenon, which precedes the
fallback accretion phase, gives rise to a unique double-peaked X-ray signature
that, if detected, may be one of the few observable diagnostics of how
stars behave under the influence of strong gravitational fields.  If
$\sim 10^{6} M_{\odot}$ black holes were prevalent in small or even
dwarf galaxies, the nearest of such flares may be detectable by EXIST
from no further away than the Virgo Cluster.
\end{abstract}

\keywords{X-rays: bursts --- shock waves --- black hole physics --- radiation mechanisms: thermal}

\section{Introduction}

\begin{figure*}[t]
\centering\includegraphics[width=0.41\linewidth,clip=true,angle=-90]{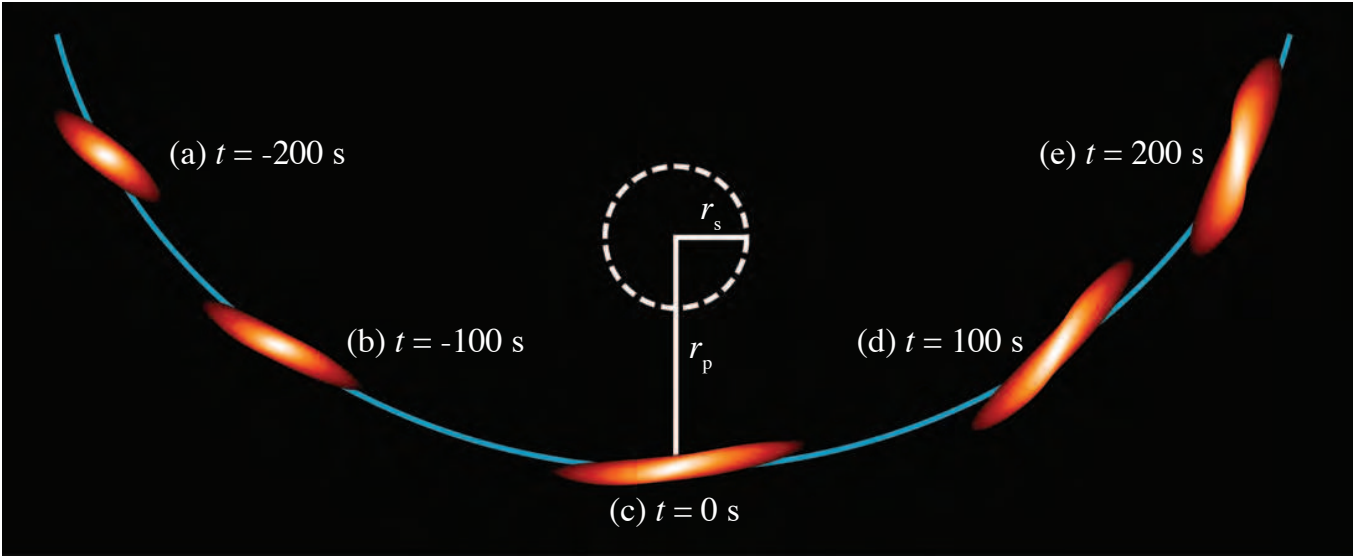}
\caption{Events in the life of a tidally-disrupted solar-type star. The star travels from left to right in the black hole's frame, with $t = 0$ s when $r = r_{\rm p}$, the pericenter passage distance. The event horizon, which lies a distance $r_{\rm s}$ from the black hole, is shown as a thick dashed line. The orange coloring represents $\log_{10} \rho$, and the light blue line shows the path of the star's center of mass. The encounter is not drawn to scale.}
\label{overheadview}
\end{figure*}

Each star in the vicinity of a supermassive black hole (SMBH) traces out a
complicated orbit under the combined influence of all the other stars
and the black hole itself. There is a chance that encounters with other
stars shift a star onto a nearly radial {\it loss cone} orbit which
brings it very close to the black hole. Exactly
how close a star can approach a black hole without suffering
distortions is defined by the tidal radius
\begin{equation}
r_{\rm t} \simeq 7\times 10^{12}\,M_{6}^{1/3} \left(R_\ast/R_\odot\right)
(M_\ast/M_\odot)^{-1/3}\;{\rm cm},
\end{equation}
where $M_{6}$ denotes the mass of the
black hole in units of $10^{6}M_\odot$. While the tidal radius is defined by average stellar properties, a solar-type star passing within this distance would likely be disrupted in a single flyby.

It is a complicated (although tractable) problem of stellar dynamics to
calculate how frequently a star enters this zone of vulnerability
\citep{Frank:1976p15}. For galaxies with steep density cusps, such
disruptions would take place about once every few thousand years
\citep{Magorrian:1999p198}. The exact rate depends on the
statistics of the stellar orbits and particularly on how quickly the
near-radial loss cone orbits are replenished. When a star is
disrupted, the sudden release
of gravitational binding energy is bound to produce a burst of radiation. The flares resulting from a disrupted
star could be the clearest diagnostic of a black hole's presence.

When a rapidly changing tidal force starts to compete with the star's
self-gravity, the material of the star responds in a complicated way
\citep{Rees:1988p9, Carter:1983p12, Bicknell:1983p604}. During a close passage, the star is
stretched along the orbital direction, squeezed at a right angle to the
orbit, and strongly compressed in the direction perpendicular to the orbital plane. This phenomenon poses a difficult
challenge to computer simulations --- three-dimensional
gas-dynamical calculations have so far addressed the fate of the bulk
of the matter, but key questions relating to the details of extreme tidal compression have yet to be answered. In
particular, very high spatial resolution is needed to model stars
passing well within the tidal radius, for which extreme compression is
halted by a shock which rebounds and eventually breaks out of the
stellar surface. In this paper, we present the most highly
resolved three-dimensional simulation of the tidal disruption of a
solar-type star to date, with voxels $\sim$$10^{3}$ times smaller
than previous calculations \citep{Lodato:2009p3051,
  Kobayashi:2004p152}. We use the results of our simulation to construct a model for the breakout of shockwaves from a tidally disrupted star, and we apply this model to determine the detectability of these tidal shock breakouts (TSBs) in the local Universe. Our results provide new insights into what
happens when stars are strongly shocked as a result of extreme
compression and the characteristic properties of such events.

The structure of the present article is as follows. A description of the numerical methods and the initial models are summarized in \S \ref{sec:nummethod}. In \S \ref{sec:hydro}, we present a chronological overview of the events in the life of a tidally disrupted star. An analytical treatment of shock breakout in tidally disrupted stars follows in \S \ref{sec:breakout}, while estimates for the luminosity and detectability of various TSB events are subsequently presented in \S \ref{sec:observability}. We identify limitations and summarize our results in \S \ref{sec:conclusion}. For convenience, a glossary of symbols (Table \ref{gloss}) is included as an appendix.

\section{Numerical Method and Initial Model}\label{sec:nummethod}
Our simulation is carried out using FLASH \citep{Fryxell:2000p440}, an
adaptive mesh code that has been used to treat a wide variety of
gas dynamics problems. The black hole is initialized as a point mass with $M_{\rm h} = 10^6 M_\odot$, while
the progenitor star is initialized as a 1 $M_\odot$, $\Gamma = 3$ polytrope \(\left(P \propto \rho^{1+1/\Gamma}\right)\) with
$R_\ast = 7 \times 10^{10}$ cm and central density of 76 g cm$^{-3}$.
During the simulation, the gas obeys a $\gamma_{\rm ad} = 5/3$ equation of state.
To ensure that the hydrostatic equilibrium of this configuration is maintained,
we ran a control simulation without the presence of the black hole for
$10^4$ s, roughly the duration of the encounter. We found good stability ---
density does not change by more than a part in 100 for the inner 99.5\%
of the star's mass.

The center of the computational domain is fixed to the star's center of mass
and is $8 \times 10^{11}$ cm on a side. We subtract out the gravitational
force the black hole exerts on the star's center of mass from every grid cell,
leaving just the tidal force, which ensures the star remains centered in
our computational domain. Self-gravity is calculated using a multipole
expansion of the star's mass distribution with only the monopole and
quadrupole moments contributing significantly.

Different refinement levels have different effective viscosities, so
it is important that the refinement of cells are chosen to match the
symmetry of the problem. We use an adaptive mesh scheme that refines
all zones with $\rho > 10^{-4}$ g cm$^{-3}$ to have 8 levels. These zones are refined twice more if they have $|z| \leq 10^{10}$ cm,
resulting in a total refinement of 10 levels near the orbital
plane where the shock is expected to form. Each block is then divided into $8^3$ grid cells, making our
smallest cells $2 \times 10^8$ cm in width. At pericenter, the star
is resolved by $10^8$ grid cells.

Because the majority of the stars in the loss cone are on radial orbits
\citep{Magorrian:1999p198}, the orbit is assumed to be parabolic. We
begin the simulation at $t = -10^4$ seconds, the star crosses the tidal radius
at $t = -10^3$, and reaches pericenter at $t = 0$. At pericenter, the star is brought to within $r_{\rm p} = 10^{12}$ cm of the black hole, which corresponds to an impact parameter $\beta \equiv r_{\rm t}/r_{\rm p} = 7$. Orbital energy is lost to heat and rotation injected into
the star during the passage, so we expect some deviation from a
parabola. However, for large $\beta$ the fractional loss of energy is
small. At pericenter, the kinetic energy of the star's bulk motion
relative to the black hole at is \(\beta G M_\odot M_{\rm h} / r_{\rm
  t} = 2 \times 10^{53}\) ergs, while our numerical results show only
\(2 \times 10^{50}\) ergs are injected as heat and internal
motions. Although general relativistic effects do affect the shape of the
orbit somewhat for $\beta$ = 7, we restrict ourselves to a purely parabolic
encounter in Newtonian gravity.

\section{Hydrodynamics of Stellar Disruption}\label{sec:hydro}

\begin{figure}[tb]
\centering\includegraphics[width=0.745\linewidth,clip=true,angle=-90]{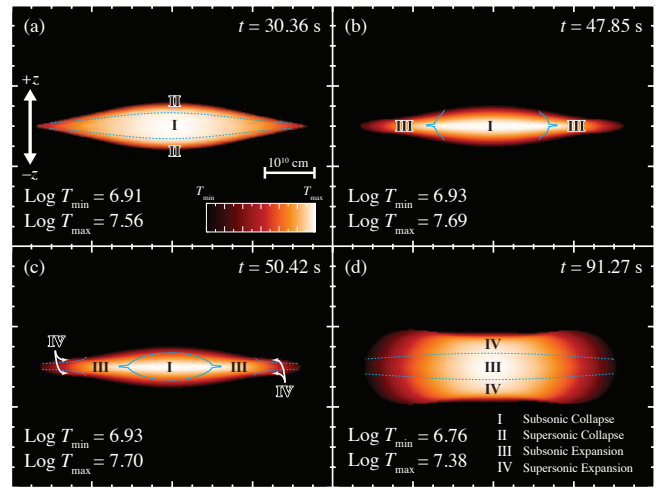}
\caption{The hydrodynamics of shock formation in the core of a tidally compressed star. The snapshots
  show the progression of gas temperature during
  the bounce phase of the passage in a slice
  that passes through the center of mass, is perpendicular to the
  orbital plane, and parallel to the minor axis of the star. The orange coloring in these
  figures indicates $\log_{10} T$ where $\rho > 1$ g cm$^{-3}$. Regions
  of subsonic/supersonic collapse/expansion are indicated with roman
  numerals, with solid contours indicating the transition from
  collapse to expansion. The dashed contour is where $M_z = v_z/c_s =
  1$, the mach 1 surface relative to the z direction. The angle of the
  minor axis to the x-z plane is $126.2^{\circ}$ in (a),
  $127.3^{\circ}$ in (b), $127.8^{\circ}$ in (c), and $141.9^{\circ}$
  in (d).}
  
\label{shocks}
\end{figure}

\subsection{Initial Approach}

A solar-type star on a loss-cone orbit is initially in hydrostatic equilibrium as it advances on a parabolic trajectory towards the black hole, eventually nearing $r_{\rm t}$, the distance at which the black hole's tidal forces are comparable to the star's self-gravity. In this region, the velocity and gravitational force vectors $\vec{v}$ and $\vec{F_{\rm g}}$ of the star relative to the black hole are almost parallel, resulting in the star being stretched into a prolate spheroid along its direction of travel (Figure \ref{overheadview}).

As the star continues in its orbit beyond $r_{\rm t}$, $\vec{F}_{\rm g}$ changes angle with respect to $\vec{v}$, and the star is no longer stretched along its major axis. The angle of the star's major axis at $r_{\rm t}$ is partially preserved during the passage. As a result, the star is not perfectly aligned with the trajectory of the center of mass. For $\beta \gg 1$, the motion of the fluid after crossing $r_{\rm t}$ (until reaching $r_{\rm p}$) is well-described by the trajectories of a system of collisionless particles. During this phase, the star is vertically compressed by the tidal field, leading to a velocity field that is directed towards the orbital plane and whose magnitude is related to $z$, the distance above the plane.

\subsection{Rebound}

In the collisionless approximation, every particle would be expected to
cross through the orbital plane shortly after the star's center of mass crossed pericenter. However, as the pressure of the gas in the orbital plane increases, it eventually becomes large enough to overcome the tidal field, which is decreasing as the star moves away from the black hole. When the vertical pressure gradient of the gas
at the leading edge of the star becomes larger than the tidal
field, the flow reverses and compression waves propagate outwards in
all directions. The compression waves
perpendicular to the orbital plane travel through a decreasing density
gradient and eventually steepen into shock waves. Shock waves that
form on the edges of the star are able to reach the surface quickly
as the star has little vertical extent in these regions. The first of these shocks form in the leading edge of the star $\sim$30
seconds after pericenter. Subsequent shock formation sweeps across the
star (parallel to the star's major axis) at $\sim v_p$, with the edges of the star rebounding before the interior, as illustrated in Figure (\ref{shocks}).

The largest densities and
pressures are reached when the compression wave crosses what was the
core of the star prior to the encounter. We find that
$P_{\rm max} = 1.3 \times 10^{18}$ dyn cm$^{-2}$ at a density $\rho_{\rm max} =
310$ g cm$^{-3}$, occurring 53.1 s after pericenter (Figure \ref{shocks}, panel c). These conditions
persist for only $\sim$10 seconds, and substantial nuclear burning
can only be triggered on this timescale for $T \sim 3\times 10^{8}$ K and $\rho \sim 10^{3}$ g cm$^{-3}$ \citep{Champagne:1992p2489}. Comparison with the results presented in \cite{Brassart:2008p455} shows that this is substantially lower than the predictions of one-dimensional simulations. This is expected because the density at the leading edge of the star decreases after infall
reversal, allowing the pressure build-up at the interior to be
relieved by leaking into the newly rarified post-bounce material. The reduction in pressure due to this effect moderates the compression waves that form
downstream in denser parts of the star by decreasing the vertical
pressure gradient $\partial_z P(x,y,t)$, a feature that is absent from one-dimensional simulations, which assume that all parts of the star collapse at the same time and ignore the effects of neighboring regions.

Additionally, a consequence of simulating tidal disruptions in three-dimensions is degraded linear resolution, which reduces the mid-plane pressure by a factor $\chi \sim l^{-1}$, where $l$ is the width of a grid cell. Applying Brassart and Luminet's scaling laws to our progenitor model, we find that they predict a compressed core pressure that is $\chi = 100$ times larger than what is found in our simulation. Because the pressure gradient and resolution effects both lead to a decrease in the mid-plane pressure, we expect that the actual value of the pressure at the mid-plane should fall somewhere between Brassart and Luminet's results and our own.

\begin{figure}[tb]
\centering\includegraphics[width=\linewidth,clip=true]{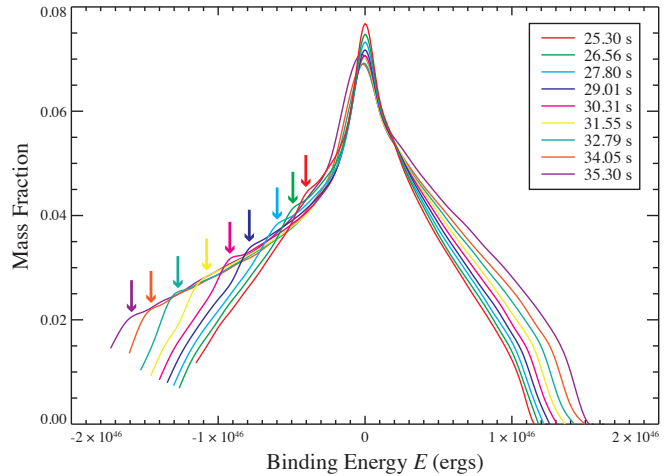}
\caption{Histogram of the energies summed over all fluid elements during
  maximum compression, $E_{i} = K_{i} - U_{i}$ where $K_{i} = \frac{1}{2} m_i
  (\vec{v}_{\mathrm{bulk}} + \vec{v}_i)^2$ and $U_{i} = G M_h m_i /
  r$. $\vec{v}_{\mathrm{bulk}}$ is the star's bulk velocity, while
  $m_i$ and $\vec{v}_i$ are the $i$th fluid element's mass and
  velocity. Energies are distributed among 30 bins and the data is
  smoothed using a spline fit. Curves are about 1.25s apart in time,
  with colored arrows indicating effect of the pressure wave
  propagating through the star.}
\label{hist}
\end{figure}

Shortly after the densest part of the star reaches maximum compression, the trailing edge enters its rebound phase. As the infalling material is
halted by internal pressure at the interior, the high pressure wave
continues to sweep through the trailing half of the star at
velocity $\sim v_p$. Because the compression wave is traveling through a
decreasing density gradient, the lighter material piles up on the
denser material towards the core, resulting in an increase of velocity
of the trailing edge relative to the star's center of mass. This
has the effect of binding more material to the black hole (Figure
\ref{hist}) because the force is opposite to the direction of the
star's motion. As a result about 5\% of the mass that was
already bound to the black hole becomes more deeply bound after the
maximum compression wave has crossed the entire
star. One-dimensional models do not replicate this effect mainly
because they assume each column of gas collapses independently,
neglecting both compression and rarefaction waves that propagate from
other regions of the star. The relative speeds
between any two parts of the star is proportional to the total spread
in velocity $\Delta v \propto v_{\mathrm{p}} \propto \beta^{1/2}$
\citep{Rees:1988p9}. Because the temperature of the star is largely
uniform and the sound speed is proportional to $\beta$
\citep{Luminet:1986p237}, a given fluid element will be causally
connected with a larger fraction of the star for larger penetration
factors, and thus multidimensional effects become increasingly important.

\subsection{Free Expansion}

\begin{figure*}[t]
\centering\includegraphics[width=0.333\textwidth,clip=true,angle=-90]{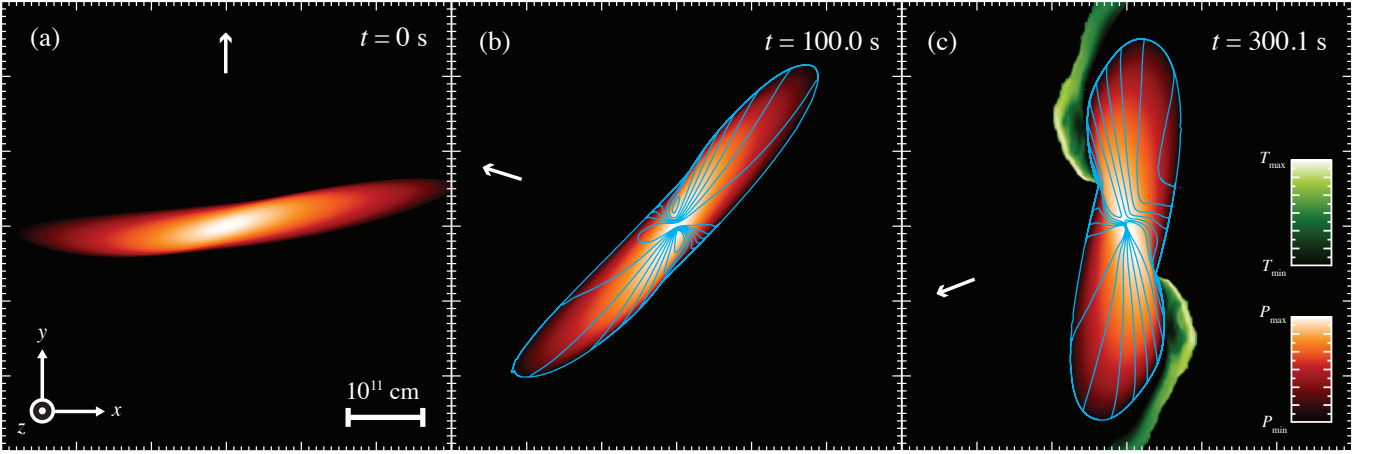}
\caption{Three snapshots showing a slice of the star in the $z = 0$
 plane, where $t$ is time after
 pericenter. The white arrow in each panel indicates the direction to
 the black hole's center. The orange coloring of the panels indicate
 $\log_{10} P$ where $\rho > 10^{-3}$, where $P_{\mathrm{min}}$ and
 $P_{\mathrm{max}}$ are set to the minimum and maximum pressure in
 each snapshot. In panels (b) and (c), contours of constant angular
 velocity $\omega = v_{xy} / |\vec{r}|$ are drawn in cyan showing
 deviations from rigid-body rotation.}
\label{difrot}
\end{figure*}

After the entire star has rebounded, it proceeds to expand and adiabatically degrade its internal energy content. Once again, the star's pressure becomes unimportant in determining its dynamics; as the star leaves the vicinity of the black hole, the effects of tidal stretching begin to dominate. As the leading edge of the star is slightly closer than it would be on a purely parabolic trajectory, the star experiences a torque and slowly rotates counter-clockwise when viewed from above (Figure \ref{difrot}).

Because different parts of the star lie at different distances to the black hole, they experience slightly different torques and acquire a range of angular velocities. This generates a spiral feature similar to that seen in
Figure 2 of \cite{Evans:1989p147}. This differential rotation leads to the star folding onto itself, and as a consequence high pressure regions in the interior are lifted towards the star's atmosphere and result in the ejection of material (third panel of Figure \ref{difrot}). These ejections becomes prominent
$\sim$200 seconds after pericenter, with the exposed surface area
being determined by the exact structure of the differential rotation
and the state of the remnant's atmosphere.

Eventually, the star leaves the immediate vicinity of the black hole and enters a free expansion phase. As approximately half of the star's mass has a negative binding energy relative to the black hole, the bound material will eventually return to pericenter \citep{RamirezRuiz:2009p3071}, form an accretion disk, and feed an AGN phase for a few months \citep{Rees:1988p9}. As we will describe in the next section, the rebound of the remnant also produces a clear observable signature, which can be used to predict when a quiescent black hole will become active.

\section{Shock Breakout}\label{sec:breakout} 
The immediate signatures of tidal disruption will be observable in
the X-rays, which correspond to a surface temperature of $\sim10^7$
K. Because the ratio between the star's vertical extent and its width in the orbital plane is less than 1:10, the light curve produced by shocks
propagating perpendicular to the orbital plane should have roughly a
$\cos(\psi)$ dependence, where $\psi$ is the angle between the orientation of the orbital plane and the line of sight. Due to the complex nature of the shocks generated by this event, the escaping radiation does not necessary lead to a single burst. In fact, our simulation shows two distinct peaks (Figure \ref{lightcurve}).

Because our simulation only resolves the star out to densities of $\sim
10^{-3}$ g cm$^{-3}$, our calculated surface lies beneath the true $\tau = 1$
surface for Thomson scattering in a stellar atmosphere. This surface corresponds to densities of $m_{\rm p} / \sigma_{\rm th} l = 10^{-8}$ g cm$^{-3}$ in our simulation, where $l$ is the length of the smallest
grid cells. Additionally, the simulation treats the entire star as a $\gamma_{\rm ad} = 5/3$ ideal gas, which is clearly not true in the tenuous outer layers of the star's atmosphere that are radiation pressure dominated for temperatures of $\sim 10^{7}$ K and densities of $\rho \lesssim 1$ g cm$^{-3}$. Because of these limitations, we must extrapolate from the conditions at the surface of the simulation to determine the properties of the post-shocked atmosphere.

We define the surface of our simulation to be where $\epsilon_{\rm gas} > 5\epsilon_{\rm rad}$ such that the gas temperature is still representative of the energy content of a given fluid element. As the photons below the $\tau = 1$ surface are trapped on the shock crossing timescale, we can assume the shock has no radiative losses and the properties of the post-shock material are accurately described by the Rankine-Hugoniot jump conditions \citep{Shu:1992p2753}. For a perfect gas of ionized Hydrogen the energy density is $\epsilon_{\rm gas} = 3nk_{\rm b} T$, where $n$ is the number density of atoms and $T$ is the pre-shock gas temperature. Thus, the post-shocked
material will have \(\epsilon = \frac{\rho}{7}\left(6 v_{\rm sh}^{2} -
k_{\rm b} T / m_{\rm p}\right)\), where $v_{\rm sh}$ is the shock velocity. This expression is controlled by the
$v_{\rm sh}^{2}$ term in the star's outer layers where $T$ is
small, and thus the energy density of a post-shocked region is $\propto \rho v_{\rm sh}^{2}$, the ram pressure of the post-shock material.

If we take the energy density $\epsilon_{\rm s} = 3\rho_{\rm s}k_{\rm b}T_{\rm s} m_{\rm p}^{-1}$ (where the subscript s refers to the surface conditions) to be equal to the post-shock energy density at the base of the atmosphere, we can then calculate the energy density $\epsilon$ as a function of height $z$ above this surface by considering the self-similar solutions of a shock propagating through a decreasing density gradient. In such solutions, the evolution of the shock velocity as it moves towards the photosphere is determined by $\gamma_{\rm ad}$ and the dependence of $\rho$ on $z$, where $z$ is the distance from the base of the atmosphere.

\subsection{Self-Similar Solutions}

Solutions for the post-shock conditions in the strong shock limit of both exponential \citep{Grover:1966p2209, Hayes:1968p2216, ZelDovich:1967p2450} and power-law \citep{Sakurai:1960p2398} atmospheres are readily available. The outer layers of an $\Gamma = 3$ polytrope are well-described by a power-law distribution with index $1.5 \le \delta \le 3$ \citep{Matzner:1999p379}; we assume that $\delta = 1.5$ for the remainder of this work. For a power-law atmosphere where $z$ is the distance from the base of the atmosphere and $h_{0}$ is the pressure scale height, the density $\rho(z)$ is
\begin{equation}
\rho(z) = \rho_{\rm s} [1 + zh_{0}^{-1}]^{-\delta}.
\end{equation}
If $v_{\rm sh,s} = \frac{7}{6} \epsilon_{\rm s}/\rho_{b}$ is the shock velocity at the base of the atmosphere, then $v(z)$ is
\begin{equation}
v(z) = \frac{2v_{\rm sh,s}}{\gamma_{\rm ad}+1} [1 + zh_{0}^{-1}]^{(1-\alpha)/\alpha}.
\end{equation}
Because the post-shock energy density is just equal to the ram pressure in the strong shock limit, the total energy density $\epsilon(z)$ is simply
\begin{equation}
\epsilon(z) = \epsilon_{\rm s} [1+ zh_{0}^{-1}]^{2(1-\alpha)/\alpha - \delta}.
\label{epsilon}
\end{equation}

At the base of the atmosphere thermal pressure dominates, but as the density of the gas decreases with $z$ more rapidly than the temperature, the atmosphere becomes radiation pressure dominated within a couple scale heights and thus $\gamma_{\rm ad} = 4/3$. For our assumed values of $\delta$ and $\gamma_{\rm ad}$, Sakurai's method leads to a similarity exponent $\alpha = 0.7774$.

As the atmosphere is strongly compressed by the time-integrated tidal gravitational field, its vertical scale at shock breakout is significantly reduced. To estimate the atmosphere's size at breakout, we assume that collapse is self-similar. Because most of the stretching in a close encounter takes place before the star reaches $r_{\rm p}$, the cross-sectional area $A_{\ast}$ should not be a strong function of $\beta$, provided that $\beta \gg 1$. Thus, the star's vertical size is directly related to the change in the star's volume
\begin{equation}
\frac{\mathcal{H}_{\rm f}}{\mathcal{H}_{\rm i}} = \frac{V_{\rm f}}{V_{\rm i}},\label{rfri}
\end{equation}
where $\mathcal{H}_{\rm i} \left(=R_{\ast}\right)$ and $\mathcal{H}_{\rm f}$ are the initial and final heights of the star above the orbital plane and $V_{\rm i}$ and $V_{\rm f}$ are the initial and final stellar volumes. The vertical velocity $v_{\perp}$ of the stellar material just prior to rebound is approximately equal to $\beta c_{\rm s}$ \citep{Carter:1983p12}, and because nearly all of the internal motions during free-fall are vertical, the total kinetic energy of the star can be estimated as
\begin{equation}
\frac{1}{2}M_{\ast}v_{\perp}^{2} = \beta^{2} \frac{G M_{\ast}^{2}}{R_{\ast}}.\label{intener}
\end{equation}

At rebound, the kinetic energy of the infalling material is converted into internal energy. Because the star's original internal energy is $U_{\rm i} \simeq G M_{\ast}^{2}/R_{\ast}$, the initial and final internal energies are simply related, $U_{\rm f} = \beta^{2} U_{\rm i}$. Assuming the compression is adiabatic,
\begin{equation}
V \propto U^{\frac{1}{1-\gamma_{\rm ad}}},
\end{equation}
and thus
\begin{equation}
\mathcal{H}_{\rm f} = \mathcal{H}_{\rm i} \left(\frac{U_{\rm i}}{U_{\rm f}}\right)^{\frac{1}{1-\gamma_{\rm ad}}} = \mathcal{H}_{\rm i} \beta^{\frac{2}{1-\gamma_{\rm ad}}}\label{hf}.
\end{equation}
For $\gamma_{\rm ad} = 5/3$, equation (\ref{hf}) gives $\mathcal{H}_{\rm f} = \mathcal{H}_{\rm i} \beta^{-3}$. In a self-similar collapse, the scale height $h_{\rm b}$ will be reduced by a factor $\mathcal{H}_{\rm f}/\mathcal{H}_{i}$, and thus $h_{\rm b} = h_{0}\beta^{-3}$.

Because the velocity of a given layer is position dependent, $\Delta v$ is non-zero and the post-shock energy density $\epsilon$ is a decreasing function of $t$. The energy output will therefore be controlled by a balance between photon diffusion and adiabatic expansion, with most of the photons being released from a layer where these two timescales are comparable. Assuming plane-parallel geometry, these timescales can be estimated for a layer at a given $z$ as
\begin{align}
\tau_{\rm ad} &\simeq \left[{\rm erf}\left(1/\sqrt{2}\right)^{-1/\gamma_{\rm ad}}-1\right]\left(\frac{\partial v(z)}{\partial z}\right)^{-1}\label{tad}\\
\tau_{\rm diff} &\simeq \frac{\tau(z)^{2} \rho(z) \sigma_{\rm th}}{c m_{\rm p}}\label{trad}.
\end{align}
where $\sigma_{\rm th}$ is the Thomson cross-section and $m_{\rm p}$ is the proton mass. Equating these two expressions and solving numerically for $z$ determines $z_{\min}$, the deepest layer in the atmosphere that will contribute significantly to the emission.

With the compressed density profile, shock jump conditions, and depth of the emitting region determined, we can calculate the luminosity and spectrum of shock breakout. As the energy density of the post-shocked material is primarily given by the energy density of radiation, we can estimate the photon temperature in the atmosphere as
\(T_{\rm ph}^{4} = \epsilon(z) / a_{\rm b}\), where $a_{\rm b}$ is the Stefan-Boltzmann constant. While the temperature does decrease as $z$ approaches $H$ (the distance from the base of the atmosphere to the photosphere), this decrease is small because $T \propto \epsilon^{1/4}$.

The spectrum of photons near the $\tau = 1$ surface may be significantly non-thermal because the photon spectrum takes a non-neglible amount of time to approach a Planckian distribution \citep{Katz:2009p2464}. Additionally, the shock velocity near the surface is $\sim0.03c$, and special relativistic effects may also contribute to a non-thermal spectrum.
While these effects would lead to the production of harder photons that are easier to detect \citep{Band:2008p586}, we conservatively assume that the observed photon spectrum is a combination of blackbodies.

The total energy released by the event can be estimated by integrating the energy density over $z$ and multiplying by twice the cross-sectional of the star at pericenter $2A_{\ast} = 2 \pi a b$:
\begin{equation}
E_{\rm tot} = 2 \pi ab \int^{H}_{z_{\min}} \epsilon(z) dz\label{toten},
\end{equation}
where $a$ and $b$ are the semi-major and semi-minor axes of the star. We can define an effective height parameter $Z$
\begin{align}
Z &\equiv \int^{H}_{z_{\min}}\left(1+ \frac{z}{h_{\rm b}}\right)^{2(1-\alpha)/\alpha - \delta}\;dz\\
&=h_{\rm b} \frac{\alpha}{\alpha(\delta+1) - 2}\left(1+\frac{z}{h_{\rm b}}\right)^{\frac{2}{\alpha}-\delta-1}\bigg\rvert^{z\;=\;H}_{z\;=\;z_{\min}},\nonumber\label{z}
\end{align}
which is independent of position on the surface of the star if we assume that the atmosphere is always perpendicular to the orbital plane and that $h_{\rm b}$ and $\delta$ are constants. The average luminosity of the event is then given by substituting $Z$ into equation (\ref{toten}) and dividing by the length of time for the star's center of mass to cross pericenter
\begin{equation}
t_{\rm cr} = 2a/v_{\rm p} \simeq \frac{6 R_{\ast}^{3/2}}{\beta^{1/2}G^{1/2}M_{\rm h}^{1/3}M_{\ast}^{1/6}},
\label{tcr}
\end{equation}
which yields
\begin{equation}
\bar{L} = \pi v_{\rm p}b\epsilon_{\rm s}Z\label{luminosityav},
\end{equation}
where we used equation (\ref{epsilon}) to substitute for $\epsilon(z)$ and our definition for $Z$. Evaluating equation (\ref{luminosityav}) for the average $\epsilon_{\rm s}$ found at the surface of our simulation, assuming the initial scale height has the solar value of $h_{0} \simeq 1.4 \times 10^{7}$ cm, and using the measured crossing time $t_{\rm cr} = 30$ s, we find $\bar{L} = 3 \times 10^{44}$ ergs s$^{-1}$.

While our estimation determines the average luminosity $\bar{L}$, calculating the luminosity as a function of time is more difficult because $\epsilon_{\rm s}$ is a function of position on the surface. Because a detailed record of the surface conditions are produced by our simulation, we can calculate what the luminosity and spectrum will look like as a function of time. Most of the luminosity is released from the deepest contributing layer of the atmosphere $\epsilon(z_{\min})$, where the diffusion time to the surface is $\sim 10^{-2}$ s. This timescale is significantly shorter than $t_{\rm cr}$, so we approximate each grid cell's contribution to the luminosity as a delta function
\begin{align}
L(x,y,t) = \frac{Z l^{2}\epsilon_{\rm s}(x,y)}{\Delta t} \delta[t - t_{\rm b}(x, y)],
\end{align}
where $l$ is the grid cell size in cm, $\Delta t$ is the time increment between two data dumps, and $t_{\rm b}(x, y)$ is the breakout time for a given location on the surface. We use $\Delta t = 0.1$s, but the value chosen only affects the smoothness of the resultant light curve so long as $\Delta t \ll t_{\rm cr}$. Because the shock front is not always easy to detect in our simulation due to limited resolution in the perpendicular direction, $t_{\rm b}$ is set by when a given area element's energy density peaks. The luminosity as a function of time is then simply a sum of the individual contributions of the surface grid cells
\begin{equation}
L(t) = \sum_{i}^{N_{\rm cells}} \frac{Z l^{2}\epsilon_{{\rm s},i}}{\Delta t} \delta(t - t_{{\rm b},i}) \label{luminosity},
\end{equation}
where $\epsilon_{{\rm s},i}$ and $t_{{\rm b},i}$ are the energy density and breakout time of a given cell, respectively.

\begin{figure}[t]
\centering\includegraphics[width=\linewidth,clip=true]{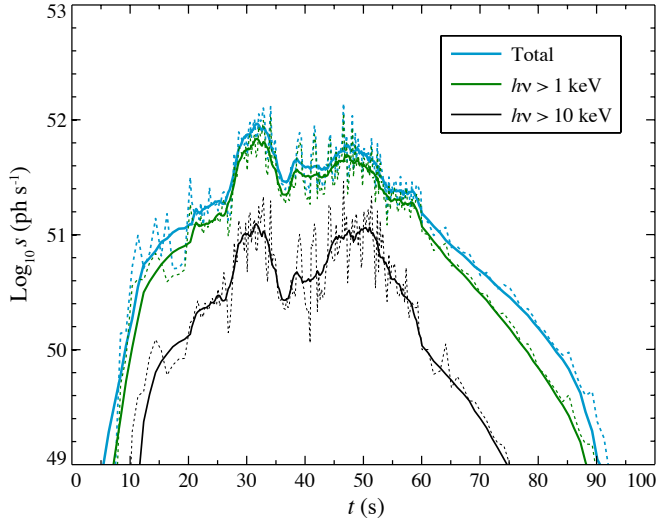}
\caption{Light curve produced by the breakout of shocks across the surface of the disrupted star, where $s$ is the flux in photons per second and $t$ is time since the star's center of mass crossed pericenter. Photons are binned into categories based on energy. The dotted lines show the raw output of the luminosity calculation for a given bin, while the solid lines show a 1 s moving average of the same data.}
\label{lightcurve}
\end{figure}

Equation (\ref{luminosity}) allows us to generate a light-curve for our disruption event (Figure \ref{lightcurve}). We assume that each grid cell produces a blackbody spectrum of photons defined by the photon temperature of that cell, $T_{{\rm ph}, i} = \epsilon_{{\rm s},i}^{1/4}a_{\rm b}^{-1}$. By summing these spectra, we can obtain a photon distribution at time $t$. Note that there are two peaks at $\sim$30 and $\sim$50 seconds corresponding to shocks that originate from the limbs of the star. As discussed in \S \ref{sec:hydro}, this double peak feature arises because compression waves that originate near the star's center of mass tend to travel parallel to the orbital plane in the direction of the star's orbital motion into regions that have already begun expanding. This compression moderation results in the non-production of shocks in this region, and thus there is a period of relatively low luminosity as the shocks cross the star's center of mass. The time difference between the two maxima in the light curve is approximately equal to $t_{\rm cr}$ (eq. \ref{tcr}). 

Our post-shock surface conditions show peak breakout temperatures in excess of $8 \times 10^{7}$ K at densities of 30 g cm$^{-3}$. \cite{Kobayashi:2004p152}
estimated that the $\beta = 5$ event will yield photons of energy of
2.2 keV on average, which is in agreement with our mean surface temperature.

\section{Observability}\label{sec:observability}

\subsection{X-ray Transient}

The duration, color, and luminosity of a disruption breakout event depends on $\beta$, $M_{\rm h}$, and the parameters of the star being disrupted. Ideally, one would want to construct a predictive model for TSBs by performing a detailed simulation for each possible combination of parameters. However, a thorough exploration of the full parameter space would require many simulations similar in scope to the simulation presented in this work. Fortunately, progress can still be made as our simulation provides an accurate benchmark for the peak luminosity $L_{\rm peak}$ of a deeply-penetrating breakout event. The results of the simulation can then be used together with some simple scaling arguments to construct a function that can describe the features of disruption breakouts for a variety of encounters.

We can gain some insight by considering the scaling of the various parameters that the luminosity (eq. \ref{luminosityav}) depends on. The pericenter velocity is dependent on the black hole mass and the closest-approach distance, which gives a scaling of $v_{\rm p} \propto \beta^{1/2}M_{\rm h}^{1/3}$. For $\gamma_{\rm ad} = 5/3$ (appropriate for the mid-plane), $\rho \propto \beta^{3}$ and $T \propto \beta^{2}$ \citep{Luminet:1986p237}, which implies that $\epsilon_{\rm s} \propto \rho v_{\rm s}^{2} \propto \beta^{5}$. Because the total height of the atmosphere $H$ scales as $\beta^{-3}$, large $\beta$ events are reduced in output by the decrease in volume of the emitting region. For ultra-close encounters the escape velocity can be comparable to $c$, and thus the emitted photons are gravitationally redshifted, with the energy of each being divided by a factor of \(\left(1 + \frac{1}{2} r_{\rm s}/r_{\rm p} \right)\). Additionally, the passage of time in the star's rest frame is slower than that of an observer for which $r \gg r_{\rm s}$ by a factor of $\sqrt{1 - r_{\rm s}/r_{\rm p}}$, leading to a further decrease in luminosity. Accounting for these effects, the peak luminosity of a TSB roughly scales as
\begin{equation}
L_{\rm peak} \propto \beta^{5/2}M_{\rm h}^{1/3}\frac{\sqrt{1 - r_{\rm s}/r_{\rm p}}}{2 + r_{\rm s}/r_{\rm p}}\label{lscaling}.
\end{equation}

\begin{figure}[t]
\centering\includegraphics[width=\linewidth,clip=true]{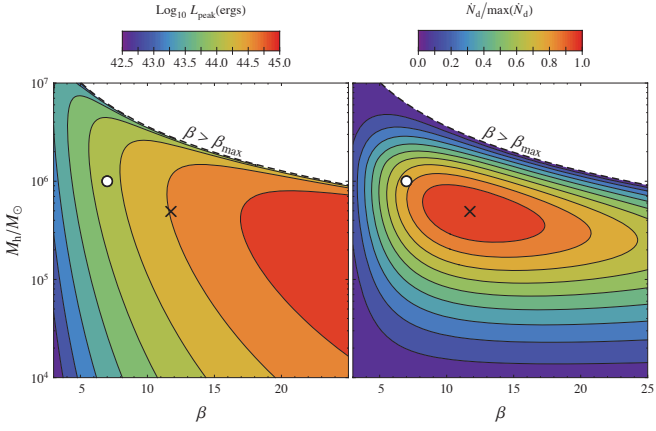}
\caption{Peak luminosity and detectability of events as a function of $M_{\rm h}$ and $\beta$, assuming that the outer atmosphere is described by a $\delta = 1.5$ power law. The left panel shows \(\log_{10} L_{\rm peak}\) for a given event, while the right panel shows the number of expected detections by EXIST \citep{Grindlay:2004p677} \(\dot{N}_{\rm d}\), normalized to the maximum rate. Our simulation parameters are indicated by the white circles, while the black crosses indicate the location of the events with the largest predicted detection rates. Disruptions are constrained by the $\beta = \beta_{\rm max}$ curve (shown as a dashed line), which indicates the largest $\beta$ for a given black hole mass where $r_{\rm p} > r_{\rm s}$.\vspace{5pt}}
\label{detectpow}
\end{figure}

A more careful calculation considering the detailed properties of the shocked, self-similar atmosphere reveals that $L$ also depends on the choice of $\delta$. The dependence on $\beta$ and $M_{\rm h}$ still roughly follows the scaling of equation (\ref{lscaling}), but because the volume of the emitting region depends on where $\tau_{\rm ad} = \tau_{\rm rad}$ (eqs. \ref{tad} and \ref{trad}), the luminosity of a given event must be computed numerically. The full numerical solution for $L_{\rm peak}$ of a $\delta = 1.5$ power-law atmosphere over a range of $\beta$ and $M_{\rm h}$ is shown in the left panel of Figure (\ref{detectpow}).

We now follow the procedure of \cite{Wang:2004p599} to estimate the rate of detectable events. Because the tidal radius grows as $M_{\rm h}^{1/3}$ and the Schwarzschild radius grows as $M_{\rm h}$, disruptions with $\beta > 3$ are only possible for black holes with $M_{\rm h} \lesssim 10^{8} M_{\odot}$. This means a significant detection rate is only obtained when considering the low end of the galaxy luminosity function, as characterized by \cite{Trentham:2002p673}
\begin{eqnarray}
N_{\rm dE}(\mathcal{M}) d\mathcal{M} &=& N_{\rm dE,0} \left(10^{-0.4(\mathcal{M}-\mathcal{M}_{\rm dE})}\right)^{\alpha_{\rm dE}+1}\label{dwn}\\
&\times&\;e^{-10^{-0.4(\mathcal{M}-\mathcal{M}_{\rm dE})}} d\mathcal{M}.\nonumber
\end{eqnarray}
In this expression, $N_{\rm dE,0}$ is a normalized dwarf elliptical galaxy number density, $\mathcal{M}_{\rm dE}$ is the cutoff magnitude for the Schechter function, and $\alpha_{\rm dE}$ describes the faint-end slope. As in \cite{Wang:2004p599}, we want to write equation (\ref{dwn}) as a function of $M_{\rm h}$. We use the scaling relation of \cite{Magorrian:1998p594} to write the absolute magnitude $\mathcal{M}$ in terms of bulge mass $M_{\rm bulge}$, which is simply related to $M_{\rm h} = 10^{-2.91} M_{\rm bulge}$ \citep{Merritt:2001p2205}. We assume that only nucleated dwarf elliptical (dEn) galaxies contain black holes, with the nucleated fraction $F_{\rm n}$ scaling linearly with $\mathcal{M}$. To properly scale $N_{\rm dE,0}$, we assume that the space-averaged dE density is equal to the number of dEs in the Virgo cluster spread into a sphere with radius equal to the distance of the Virgo cluster (approx. 0.2 Mpc$^{-3}$).

Because we are in the regime where $r_{\rm p} \ll r_{\rm t}$, we know that the disruption rate scales linearly with $r_{\rm p}$ \citep{Rees:1988p9}, and Wang's expression for the disruption rate becomes
\begin{equation}
\dot{N} = 6.5 \times 10^{-4}\;{\rm yr}^{-1}  \left(\frac{M_{\rm h}}{10^{6}\;M_{\odot}}\right)^{-0.25} \beta^{-1}.\label{wangdis}
\end{equation}

The maximum distance $R_{\rm d}$ to a detectable event can be expressed as a function of instrumental sensitivity as
\begin{equation}
R_{\rm d} = \sqrt{\frac{\pi L_{\rm peak}}{2\sigma_{\rm b}T_{\rm ph}^{4}}\int_{\nu_{\min}}^{\nu_{\max}}\frac{F_{\rm T}^{-1}B_{\nu}(T_{\rm ph})}{h\nu}\;d\nu},
\end{equation}
in which $F_{\rm T}(\nu)$ is the burst sensitivity as a function of frequency, the TSB spectrum is given by the Planck function $B_{\nu}$ at a temperature $T_{\rm ph} \propto \beta^{5/4}$, $h$ is Planck's constant, $\sigma_{\rm b}$ is the Stefan-Boltzmann constant, and $\nu_{\min}$ and $\nu_{\max}$ are the minimum and maximum frequencies accessible to the instrument. An instrument is sensitive to all events contained within the volume
\begin{equation}
V_{\rm d} = \frac{4}{3}\pi R_{\rm d}^{3}.
\label{vd}
\end{equation}

\begin{figure}[t]
\centering\includegraphics[width=\linewidth,clip=true]{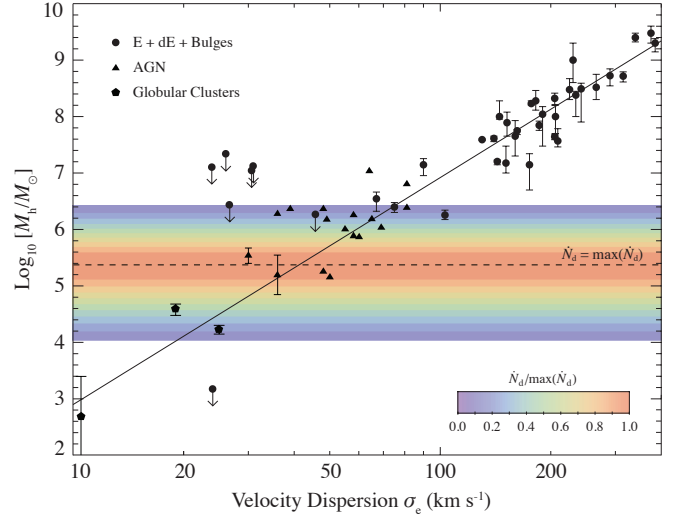}
\caption{Bulge velocity dispersion $\sigma_{\rm e}$ vs. BH masses in E, dE, and bulge-dominated spiral galaxies (circles), AGN (triangles), and globular clusters (pentagons). The colored contours represent the detection rate relative to the maximum detection rate as in Fig. (\ref{detectpow}), averaged over all $\beta$, with the maximum indicated by the dashed line. Plot modified from \citet{Geha:2002p2490}, including the BH masses compiled by \cite{Noyola:2008p2744}.} 
\label{msigma}
\end{figure}

We now have all the pieces needed to determine a detection rate in terms of $\beta$ and $M_{\rm h}$. By multiplying the number density of black holes $N_{\rm dE}F_{\rm n}$ (eq. \ref{dwn}) by the disruption rate $\dot{N}$ for a black hole of a given mass and penetration factor (eq. \ref{wangdis}), and by the volume $V_{\rm d}$ in which those events are detectable (eq. \ref{vd}), we obtain the following expression for the total rate of detection
\begin{equation}
\dot{N}_{\mathrm{d}} = \frac{\Omega}{4\pi}\int_{3}^{\infty}\int_{M_{\min}}^{M_{\max}}\dot{N}N_{\rm dE}F_{\rm n}V_{\rm d}\;dM_{\rm h} d\beta,\label{detectrate}
\end{equation}
in which $\Omega$ is the solid angle covered by the instrument's field of view and \(M_{\max} \equiv \left( c^{2} R_{\odot} / 4 G M_{\odot} \beta
\right)^{3/2}\) is the mass at which $r_{\rm p} \leq r_{\rm s}$, the Schwarzschild radius. The integral over $\beta$ terminates at $\beta = 3$ because shocks have not been seen in one-dimensional calculations for passages with $\beta < 3$ \citep{Brassart:2008p455}. However, our simulation shows that shocks form in the limbs of the star where conditions are substantially different than the core, and thus shocks may still be produced for $\beta < 3$ in some cases.

The detection rate is evaluated using the parameters of the proposed EXIST telescope, which has spectral coverage from 3 to 1000 keV, an average burst sensitivity of $F_{\rm T} = 0.2$ cm$^{-2}$ s$^{-1}$, and $\sim20\%$ sky coverage \citep{Grindlay:2004p677, Band:2008p586}. Using these values, equation (\ref{detectrate}) predicts that EXIST should detect approximately 1 TSB per year. We can also use the argument of integrals of this equation to generate a map in ($\beta, M_{\rm h}$) space (Figure \ref{detectpow}, right panel) to predict which events will generate the most detections. Comparison with the left panel of the same figure shows that these events have a characteristic luminosity of $\sim 5 \times 10^{44}$ ergs. Despite being similar to the Eddington value of $10^{44} M_{6}$ for a SMBH, detection is infrequent because of the short duration of the shock breakout. By integrating equation (\ref{detectrate}) over all $\beta$, we can also determine how black holes of different masses contribute to the detection rate (Figure \ref{msigma}), with the peak rate corresponding to a black hole mass of $2 \times 10^{5} M_{\odot}$. The predicted peak in the detection rate is predicated on the assumption that SMBHs obey the black hole to bulge mass relation for $M_{\rm h} < 10^{6} M_{\odot}$. Therefore, the detection of TSBs would test the validity of this assumption and potentially provide compelling evidence for SMBHs in low mass galaxies.

\subsection{Gravitational Wave Signal}

In addition to the X-ray breakout signature, gravitational waves are
also expected to radiate from the encounter, mostly originating from
the changing location of the star relative to the black hole. We can
estimate the strength of these waves by approximating the second
time derivative of the moment of inertia tensor as \(\ddot{I}\,^{ij} \sim G M R^2 /c^4
P^2\). If we assume that the period \(P \sim r_{\rm p}/v_{\rm p} = r_p^3 / G M_{\rm h}\), the length scale of variation $R \sim r_{\rm p}$, and the mass
$M = M_\ast$, we find that the gravitational wave amplitude $\bar{h}$ is \citep{Kobayashi:2004p152}
\begin{equation}
\bar{h} \sim \frac{G M_\ast r_{\rm s}}{d c^{2} r_{\rm p}} = \frac{\beta G^2 M_\ast^{4/3} M_{\rm h}^{2/3}}{d c^4 R_\ast},
\label{gravh}
\end{equation}
in which $d$ is the distance to the event. For our $\beta$ = 7 simulation, we
calculate that $\bar{h} \sim 10^{-21}$ for $d$ = 10 Mpc. Gravitational waves
will also be radiated as the star itself changes shape, but these
distortions are far smaller than those generating by the changing quadrupolar moment of the star-SMBH system. Assuming that $R \sim R_\ast$, the gravitational wave amplitude is
\begin{equation}
\bar{h} \sim \frac{\beta^3 G^2 M_\ast^2}{d c^4 R_\ast}.
\end{equation}
Note that this expression is independent of $M_{\rm h}$. For $d$ = 10 Mpc, $\bar{h}$ is $10^{-23}$ for our encounter, beyond LISA's sensitivity range. Because this expression is proportional to $\beta^{3}$, the compression of the star can be a substantial contribution to the gravitational wave signature for very deep passages ($\beta \gtrsim 25$).

If both a gravitational wave signal and a TSB signal are available for the same event, additional information about the encounter can be obtained. While both observational signatures each place upper and lower limits on the properties of the disrupted star, either signature on its own cannot uniquely constrain the star's characteristics. By assuming a stellar mass-radius relationship $M_{\ast} = M_{0}R_{\ast}^{\eta}$, equation (\ref{gravh}) can be combined with equation (\ref{tcr}) to calculate the mass of the disrupted star
\begin{equation}
M_{\ast} \sim \left(\frac{\bar{h}dc^{4}t_{\rm cr}^{2}}{GM_{0}^{2\eta}}\right)^{\frac{1}{1 - 2\eta}},
\end{equation}
in which $t_{\rm cr}$ can be estimated by measuring the distance between the two peaks in the light curve (e.g. Figure \ref{lightcurve}). By determining the masses of disrupted stars, a distribution of stars that occupy the loss cone can be derived. As the IMF in the vicinity of SMBHs is poorly characterized even in our own galaxy \citep{Alexander:2005p606}, a coincident detection of both the gravitational and TSB signatures would, for the first time, allow us to investigate the IMF in close proximity to extragalactic SMBHs.

\section{Conclusion}\label{sec:conclusion}
While the simulation presented in this work features the highest resolution
of the tidal disruption of star in 3D to date, it still has a few shortcomings. Our model does not have enough linear resolution to resolve the sharp pressure gradients that develop in deeply penetrating encounters, and thus the mid-plane pressure in the rebound phase is certainly underestimated by some factor. In addition, for close passages such as ours, GR
effects start to become important. For $5 \lesssim \beta \lesssim 10$, the orbit is better characterized by a Paczynski-Wiita potential, but for $\beta > 10$, a fully general relativistic treatment
of the Schwarzschild metric \citep{Frolov:1994p2} is required. The orbits are then not ellipses, but may have two
or more pericenter transversals and, as a result, could lead to the
formation of multiple shocks. A full understanding of the compression
process in such cases will require detailed GR hydrodynamical
simulations.

In this work, we have provided numerical details of how a solar-type star is stretched,
squeezed, and strongly shocked during an encounter with a massive black
hole. We then calculated the radiation a distant observer might detect as
the observational signature of the accompanying shock breakout. If detected, an $L \sim 10^{44}$ erg/s burst that fades within a few minutes and exhibits the predicted double-peaked signature in the soft X-rays would be compelling testimony that a star experiencing an ultra-close encounter with a black hole can be disrupted and compressed to such an extent that shock waves can be triggered.

\acknowledgments We have benefited from many useful discussions with D. Fox, J. Grindlay, C. Matzner, P. Meszaros, T. Plewa, M. Rees and L. Roberts. We thank the referee 
and R. Bernstein for useful comments. The
software used in this work was in part developed by the DOE-supported
ASCI/Alliance Center for Astrophysical Thermonuclear Flashes at the
University of Chicago. Computations were performed on the Pleaides
UCSC computer clusters. This work is supported by NSF: PHY-0503584
(ER-R) and DOE SciDAC: DE-FC02-01ER41176 (JG and ER-R).

\bibliographystyle{apj}
\bibliography{apj-jour,2008a}

\clearpage
\appendix

\begin{table*}[h]
\centering
\caption{Glossary of symbols.}
\begin{tabular}{ll@{\hspace{20pt}}ll}
\hline
 Term & Meaning & Term & Meaning\\
\hline
\it{Fundamental constants} &&									\it{Shock breakout}\\
$a_{\rm b}$ & Radiation constant &								$h_{0}$ & Initial atmospheric scale height\\
$c$ & Speed of light &										$h_{\rm b}$ & Atmospheric scale height at shock-breakout\\
$G$ & Newton's constant &									$H$ & Depth of compressed atmosphere\\
$h$ & Planck's constant &										$n$ & Atomic density\\
$k_{\rm b}$ & Boltzmann's constant &							$t_{\rm b}$ & Shock-breakout time\\
$m_{\rm p}$ & Proton mass &									$T$ & Gas temperature\\			
$M_{\odot}$ & Solar mass &									$T_{\rm ph}$ & Photon temperature\\
$R_{\odot}$ & Solar radius &									$v$ & Post-shock velocity of atmosphere\\
$\sigma_{\rm b}$ & Stefan-Boltzmann constant &					$v_{\rm sh}$ & Shock velocity\\	
$\sigma_{\rm th}$ & Thomson cross-section &						$z_{\min}$ & Depth of deepest emitting layer\\
&&														$Z$ & Effective height\\	
\it{Encounter parameters} &&									$\alpha$ & Similarity exponent\\
$a$ & Semi-major axis of elongated star &						$\gamma_{\rm ad}$ & Adiabatic index\\
$A_{\ast}$ & Cross-sectional area of elongated star & 				$\delta$ & Power-law index of stellar atmosphere\\
$b$ & Semi-minor axis of elongated star &						$\epsilon$ & Total energy density\\
$\mathcal{H}$ & Star's extent above orbital plane&					$\epsilon_{\rm gas}$ & Ideal gas energy density\\
$M_{\rm h}$ & Black hole mass &								$\epsilon_{\rm rad}$ & Radiation energy density\\
$M_{6}$ & Black hole mass in $10^{6} M_{\odot}$ &					$\rho$ & Mass density\\
$M_{\ast}$ & Mass of star &									$\tau$ & Optical depth\\
$M_{0}$ & Stellar mass-radius normalization &					$\tau_{\rm ad}$ & Atmosphere expansion timescale\\
$r_{\rm s}$ & Schwarzschild radius &							$\tau_{\rm rad}$ & Radiation diffusion timescale\\
$r_{\rm t}$ & Tidal radius &									\\
$r_{\rm p}$ & Distance at pericenter &							\it{SMBH distribution}\\									
$R_{\ast}$ & Initial stellar radius &								$F_{\rm n}$ & Nucleated fraction of dE galaxies\\				
$t$ & Time relative to time of pericenter &							$M_{\rm bulge}$ & dE bulge mass\\							
$t_{\rm cr}$ & Time for star to cross pericenter &					$\mathcal{M}$ & Galaxy magnitude\\						
$U$ & Total internal energy of star&								$\mathcal{M}_{\rm dE}$ & dE Schechter function cutoff magnitude\\	
$v_{\rm p}$ & Velocity of star at pericenter &						$N_{\rm dE}$ & dE number density\\	
$v_{\perp}$ & Velocity of star's vertical collapse &						$\alpha_{\rm dE}$ & dE Schechter function faint-end slope\\		
$V$ & Volume of star&										\\
$\beta$ & Penetration factor &									\it{Observability}\\	
$\Gamma$ & Polytropic index of star &							$B_{\nu}$ & Planck distribution\\
$\eta$ & Stellar mass-radius relation exponent &					$d$ & Distance of disruption from observer\\
&&														$\bar{h}$ & Gravitational wave amplitude\\
\it{Simulation parameters}&&									$F_{\rm T}$ & Burst sensitivity of instrument (ph s$^{-1}$ cm$^{-2}$)\\
$l$ & Width of smallest grid cells&								$L$ & Bolometric luminosity\\
$P_{\max}$ & Pressure maximum in simulation volume&				$\bar{L}$ & Average bolometric luminosity\\
$T_{\rm s}$ & Gas temperature at simulation surface&				$L_{\rm peak}$ & Peak bolometric luminosity\\
$\Delta t$ & Time between simulation plot files&					$\dot{N}$ & Tidal disruption rate of a given SMBH\\
$v_{\rm s}$ & Post-shock velocity at simulation surface&				$R_{\rm d}$ & Maximum detectable distance\\
$v_{\rm sh,s}$ & Shock velocity at simulation surface&				$V_{\rm d}$ & Volume of sphere with radius $R_{\rm d}$\\	
$z$ & Distance from orbital plane&								$\Omega$ & Solid angle coverage of instrument\\				
$\epsilon_{\rm s}$ & Total energy density at simulation surface&		\\
$\rho_{\max}$ & Mass density maximum in simulation volume&		\\									
$\rho_{\rm s}$ & Mass density at simulation surface&				\\
$\chi$ & Pressure reduction due to limited resolution&				\\
\hline
\end{tabular}
\label{gloss}
\end{table*}

\clearpage

\end{document}